\documentclass[aps,prb,showpacs,amsmath,twocolumn]{revtex4}

\usepackage{amsmath}
\usepackage{amssymb}
\usepackage{graphicx}

\newcommand{\vareps}{\varepsilon}

\begin{document}

\title{Andreev reflection through Fano resonances in molecular wires}

\author{A.~Korm\'anyos, I.~Grace, and C.~J. Lambert}
\affiliation{Department of Physics, Lancaster University,
Lancaster, LA1 4YB, UK}


\begin{abstract}
We study Andreev reflection in a  
normal conductor-molecule-superconductor junction using a first principles 
approach. In particular, we focus on a family of molecules 
consisting  of a molecular backbone 
and a weakly coupled side group.   We show that the presence of 
the side group can lead to a Fano resonance in the Andreev reflection. 
We use a simple theoretical model to explain the results of the numerical 
calculations and to make predictions about the possible sub-gap  
resonance structures in the Andreev reflection coefficient. 

\end{abstract}


\maketitle


Fano resonances\cite{ref:fano} are a  
 universal interference phenomenon which can affect  
coherent electrical transport through nanostructures in
many different systems. 
Examples of  Fano lineshape in mesoscopic systems 
include scanning tunnelling microscope measurements on a single
 magnetic atom absorbed on a gold surface\cite{ref:madhavan,ref:schneider},
single-electron transistors fabricated into a gated
two-dimensional electron gas\cite{ref:gores}, 
quantum dots embedded into an 
Aharonov-Bohm ring\cite{ref:kobayashi-2002,ref:kobayashi-2003},
multiwall carbon  nanotubes\cite{ref:kim,ref:yi,ref:chandrasekhar}
and recently  single-wall carbon nanotubes\cite{ref:babic} 
and double-wall nanotubes\cite{ref:iain}.
Fano resonances (FRs)  also appear  in the the conductance of 
quasi one-dimensional quantum wires with donor 
impurities\cite{ref:tekman} and in the case of  
quantum wires with a  side coupled quantum dot\cite{ref:franco}. 
In molecular electronics, due to the realistic treatment of 
the metal electrodes, FRs have been found in the 
transmission  of dithiol benzene\cite{ref:grigoriev}. More 
generally,  theoretical calculations  predict 
that Fano-lineshape should appear in the transmission through 
molecular wires with attached side groups\cite{ref:thodoris}
or as a consequence of quantum interference  between surface
states of the measuring electrodes and the molecular 
orbitals\cite{ref:shi}.

If one of the measuring probes is superconducting, the conductance  
for energies $E$ smaller than the superconducting pair potential
$\Delta$ depends on the Andreev reflection probability $R_A(E)$.
The Andreev reflection in various mesoscopic systems has been
studied for a long time 
(see e.g Refs.~\onlinecite{ref:beenakker-2,ref:raimondi} 
and references therein) but the interest has recently
renewed when Andreev reflection through  carbon nanotubes 
was measured experimentally\cite{ref:morpurgo,ref:graeber}. These
experiments have sparked numerous theoretical studies 
both in the absence of the electron-electron 
interaction\cite{ref:wei,ref:pan} and in the presence 
of the interaction\cite{ref:vishveshwara,ref:titov,ref:cuevas,ref:schwab,
ref:clerk, ref:sun,ref:avishai,ref:splett,ref:doma,ref:domanski}. 
In many of  these studies it was assumed that it was sufficient to 
consider resonant transport through a single energy level and as a 
consequence, the Andreev reflection as a function of energy 
exhibited  Breit-Wigner type resonances. A notable exception is 
 Ref.~\onlinecite{ref:golub} where transport through an Aharonov-Bohm
ring with an interacting quantum dot situated in one of its arms 
 was considered and a Fano-type asymmetric resonance was found in the 
conductance. Very recently, Tanaka \emph{et al.}\cite{ref:tanaka} 
studied  Andreev transport through side-coupled interacting 
quantum dots focusing on the interplay of Andreev scattering 
and Kondo effect.  

It was demonstrated in Ref.~\onlinecite{ref:thodoris} 
that Fano resonances  are a generic feature 
of molecular wires with attached side groups. 
It was also shown that for a certain type of molecular wires a FR 
can appear in the normal conductance $G_N(E)$ very close to the 
Fermi energy $E_F$.  In a normal metal-molecule-superconductor 
(N-Mol-S) junction therefore these FRs would also affect the 
sub-gap transport. 
The aim of this paper is to study  Andreev reflection 
through  molecular wires when the normal conductance exhibits
FRs close to the Fermi energy.  Performing 
\emph{ab initio} simulations of  molecular wires in 
N-Mol-S junctions we show how FRs influence the sub-gap transport. 
We elucidate the results
of the numerical calculations using a simple analytic model.
We also predict that for finite energies 
the differential conductance can reach the unitary limit 
 if there is a strong 
asymmetry in the coupling to the leads. 

\begin{figure}[hbt]
\includegraphics[scale=0.5]{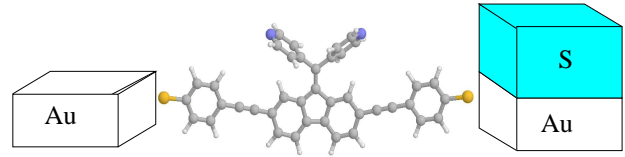}
\caption{Possible experimental setup. The molecule is contacted by
gold electrodes, one of which is superconducting due to the
proximity effect. 
\label{fig:exp-setup}}
\end{figure}
A possible experimental setup to measure  Andreev reflection in
 N-Mol-S junctions is shown in Fig.~\ref{fig:exp-setup}. 
The molecule is contacted with thin gold electrodes on both sides. 
On top of one of the electrodes a second layer of e.g. aluminium or
niobium is deposited, which at low enough temperature  
becomes superconducting.  Due to 
the proximity effect this top layer  induces 
superconductivity in the gold electrode beneath 
(in our calculations we assume that the induced superconductivity 
is $s$ type). We note that this setup was  
successfully used in Ref.~\onlinecite{ref:graeber} to study 
Andreev-reflecion in normal 
conductor - carbon nanotube - superconductor (N-Cn-S) junctions.

To study  FRs  in a N-Mol-S system, we choose 
the smallest molecule of a recently synthesized family of
molecular wires\cite{ref:wang1,ref:wang2,ref:wang3}. 
Since these molecules have terminal thiol groups they can 
easily bind to gold surfaces  making them ideal for 
experiments on single-molecule transport properties. 
The central part of the molecule consists of a single fluorenone 
unit, which could be chemically modified, e.g. by replacing 
the oxygen with  bipyridine rings, as  shown schematically in 
Fig.~\ref{fig:exp-setup}. 
The differential conductance  of the system was calculated  using 
a combination of the DFT code SIESTA\cite{ref:siesta} and a Green's 
function scattering approach explained in Refs.~\onlinecite{ref:sanvito,ref:smeagol}. 
Initially the isolated molecule is relaxed to find the optimum geometry,
then the molecule is extended  to include surface layers 
of the gold leads.
In this way,  charge transfer at the gold-molecule interface is included
self-consistently. The number $N_g$ of gold layers is
increased until computed transport properties between the 
(normal conducting) gold leads no longer changed with increasing $N_g$.
Typically, this extended molecule contained $N_g=3$ to $4$ 
gold layers on each side, 
and the layers consisted of  9 atoms on the (111) plane. 
The leads, which were assumed to be  periodic in the transport direction, 
also consisted of gold layers containing  9 atoms on the (111) plain.
Using a double--$\zeta$ basis plus polarization orbitals,
Troullier--Martins pseudopotentials\cite{ref:pseudopot} and the
Ceperley--Alder LDA method to describe the exchange correlation
\cite{ref:zunger},  effective tight-binding Hamiltonians $H_{M}$, $H_{L}$ 
of the extended molecule and of the leads, respectively,  were obtained.
To investigate the generic physics of this system, we employ the 
simplest possible approximation for the order parameter, namely 
that it changes in a step-function-like manner at the superconducting lead -
extended molecule interface. Therefore 
the superconducting lead was modelled by 
introducing  couplings of constant magnitude $\Delta$ between 
the electron and hole degrees of freedom in $H_L$, 
while no such coupling was present in the extended molecule and  
in the normal lead. 
We focus on  sub-gap transport and therefore compute 
the Andreev-reflection probability $R_A(E)$, 
because at zero temperature for $E<\Delta$ 
the differential conductance is  
given by $G_{NS}(E)=\frac{4 e^2}{h} R_A(E)$.

The calculations of Ref.~\onlinecite{ref:thodoris} have shown that 
by changing by rotational conformation of the bipyridine unit 
it is possible to change the position of the Fano resonance with 
respect to $E_F$. The definition of the  angle of rotation $\theta$ 
of the bipyridine group is  the following: $\theta=0^{\circ}$ 
when the rings of the sidegroup are parallel to the molecule axis
and it is $90^{\circ}$ when they lie perpendicular. 

\begin{figure}[hbt]
\includegraphics[scale=0.5]{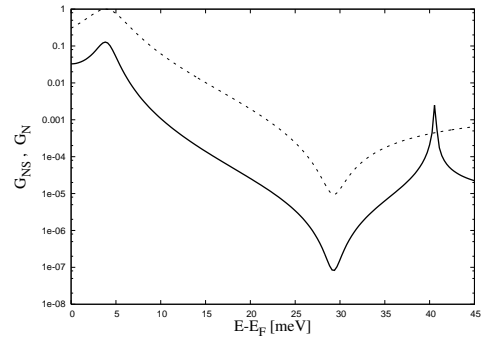}
\caption{Normal differential conductance 
$G_N(E)$ (in units of $2 e^2/h$, dashed line) 
and sub-gap conductance $G_{NS}(E)$ 
(in units of $4 e^2/h$, solid line) 
in logarithmic scale as a function of energy. We used $\Delta=41$ meV. 
\label{fig:trans2}}
\end{figure}

\begin{figure}[hbt]
\includegraphics[scale=0.5]{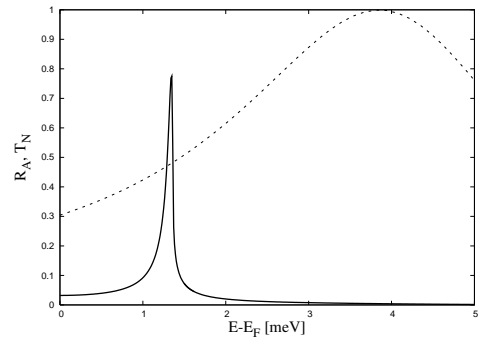}
\caption{$T_N(E)$ (dashed line) and $R_{A}(E)$ 
(solid line) as a function of energy. We used $\Delta=1.35$ meV. 
The sharp peak in $R_A(E)$ is at $E\approx \Delta$ (see main text).
\label{fig:trans1}}
\end{figure}

We consider the  molecule whose rotational conformation is 
$\theta=71.4^{\circ}$.  Assuming first that both leads are 
normal conducting  (N-Mol-N junction), close to the Fermi energy 
there is  a FR in the differential conductance 
$G_N(E)=\frac{2 e^2}{h} T_N(E)$ (where $T_N(E)$ is the normal transmission) 
as it can be seen in Fig.~\ref{fig:trans2}. 
Since for conventional supercondcutors 
the typical superconducting gap values 
are $0.1-1.5$\,meV, we first consider the transport for
energies $E<\Delta=1.35$ meV. 
The Fano peak in $G_N(E)$ is at $\delta E\approx 3.7$\,meV 
above  $E_F$, therefore $\delta E$ is bigger than $\Delta$ and  the
 influence of this resonance on the sub-gap transport 
can be understood by considering the zero bias conductance $G_{NS}(0)$. 
Indeed, as Fig.~\ref{fig:trans1} shows,  the Andreev reflection
is almost constant apart from the region $E\approx \Delta$ where  
a sharp peak can be observed which is due to the singularity in the
 density of states of the superconductor at this energy.  
One can see that off-resonance $G_{NS}(0)$ is smaller than $G_N(0)$.

More generally however, if there is a narrow  
$\Gamma \lesssim \Delta$ resonance at some 
$|\delta E| < \Delta$  above or below $E_F$,  
the energy dependence 
of the Andreev reflection becomes important. 
(In case e.g. of carbon nanotubes, which can be gated, 
this scenario should be easily attainable, as 
in Ref~\onlinecite{ref:babic} where the width of the Fano peak
was $\approx 0.2$\,meV.)
To illustrate 
this case, we performed computations using the same molecule 
but  much bigger  $\Delta$. 
The results of the computations are shown in Fig.~\ref{fig:trans2}.
As one can see, a Fano resonance now appears  both $G_{N}$ 
in $G_{NS}$. However, 
a closer inspection reveals that the width of the
Fano peak in $G_{NS}$ at $\approx 3.7$ meV is roughly half of the 
width of the   corresponding peak in $G_N$.

To explain the results of the numerical calculations we 
consider a simple model, introduced in Ref.~\onlinecite{ref:thodoris}, which
was shown to capture the essential features of the transport between normal
conducting leads. 
Close to a resonance, it is sufficient to consider a single backbone 
state  $|f_1\rangle$ with resonant energy $\tilde{\vareps}_1$ 
and a state $|f_2\rangle$ of energy $\tilde{\vareps}_2$
 which is associated with  a side group of the molecule 
($\tilde{\vareps}_1$ and $\tilde{\vareps}_2$ are measured relative to 
the  $E_F$).
The weak coupling between the backbone of the molecule and the side group is
described by a matrix $H_{12}$.   We denote by 
 $t_c=\langle f_2|H_{12}|f_1\rangle$
the coupling between the two states, whereas the coupling of the 
backbone state to the normal (superconductor) lead is
described by matrices $W_N$ ($W_S$). 
 A brief derivation of the Andreev reflection probability
$R_A(E)$ for this system is given in Appendix \ref{sec:r_a-deriv}, 
here we only summarize the main results.

The linear conductance is given by
\begin{equation}
G_{NS}(0)=\frac{4e^2}{h} 
\frac{4 \Gamma_L^2\Gamma_R^2 \tilde{\vareps}_2^4}
{[(\tilde{\vareps}_{+}\tilde{\vareps}_{-})^2+(\Gamma_L^2+\Gamma_R^2)
\tilde{\vareps}_2^2]^2}.
\label{eq:gns-zero}
\end{equation}
Here $\Gamma_L$, ($\Gamma_R$) is 
the normal state tunnelling rate to the left (right) lead at $E_F$ and
$\tilde{\vareps}_{\pm}=\bar{\vareps}\pm \sqrt{\delta\vareps^2+t^2}$
where $\bar{\vareps}=(\tilde{\vareps}_1+\tilde{\vareps}_2)/2$,
$\delta\vareps=(\tilde{\vareps}_1-\tilde{\vareps}_2)/2$.
The maximal conductance is attained at $\Gamma_L=\Gamma_R$, 
$\tilde{\vareps}_{\pm}=0$  when it is twice as 
large as the normal conductance. Note that the conductance maximum is 
not attained when  $\tilde{\vareps}_1$ is aligned with 
$E_F$ as one might expect. The hybridization between $\tilde{\vareps}_1$ 
and $\tilde{\vareps}_2$ due to the coupling $t_c$ leads 
to a different resonance condition for this system. Off-resonance, 
i.e. when $\tilde{\vareps}_{\pm}\neq 0$ ,  $G_{NS}$ falls off more
rapidly as a function of $\tilde{\vareps}_{\pm}$ than $G_{N}$ 
[see Eq.~(1) in Ref.~\onlinecite{ref:thodoris}].  Therefore the
$G_{NS}(0)$ is usually smaller than $G_{N}(0)$. 
Moreover, $G_{NS}$  is zero if   $\tilde{\vareps}_2=0$, i.e. when the 
energy of the side coupled state equals $E_F$.

\begin{figure}[hbt]
\includegraphics[scale=0.7]{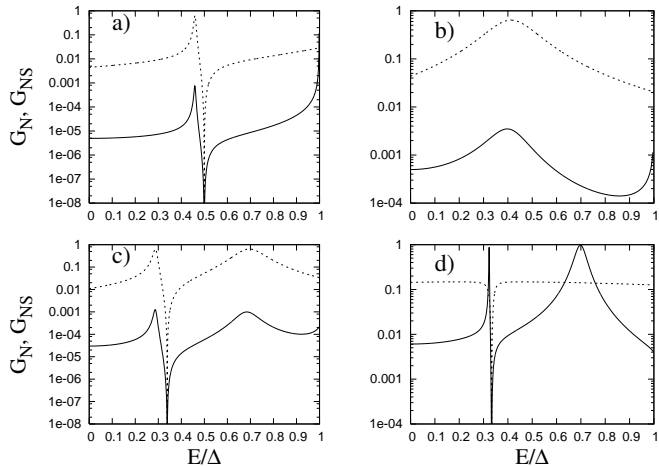}
\caption{$G_{N}$ (in units of $2e^2/h$, dashed)  and 
$G_{NS}$ (in units of $4e^2/h$, solid) in 
logarithmic scale as a function of energy. 
We used $\sqrt{\Gamma_N^e \Gamma_N^h}/|\sigma_S^{eh}|=4$  
 in the case of a), b), c)  
and $\sqrt{\Gamma_N^e \Gamma_N^h}/|\sigma_S^{eh}|=0.25$,
in the case of d).
\label{fig:theorycurves}}
\end{figure}
For finite energies $E<\Delta$ the most important features of the 
differential conductance of our model are the following. 
If the coupling to the normal lead is stronger 
than to the superconducting one, i.e. when 
$\sqrt{\Gamma_N^e \Gamma_N^h}\gtrsim |\sigma_S^{eh}|$ where
$\Gamma_N^e$ ($\Gamma_N^h$) are tunnelling 
rates for electrons (holes) from the normal lead and  
$\sigma_S^{eh}$ is an off-diagonal 
element of the self-energy matrix $\Sigma_S$ 
(see Appendix \ref{sec:r_a-deriv} for the precise definitions, as well as
for the definitions of ${\vareps}_+^e$, ${\vareps}_-^e$, 
to be introduced below), in good approximation 
\begin{equation}
G_{NS}(E)=\frac{4e^2}{h} A(E)\, T_e(E)
\label{eq:gns-finitE}
\end{equation}
where the amplitude $A(E)$ is a slowly varying function of the energy and
 \begin{equation}
 T_e(E)=
\frac{\Gamma_N^e \Gamma_N^h (E-\tilde{\vareps}_2)^2}
{[(E-{\vareps}_+^e)(E-{\vareps}_-^e)]^2+
(\Gamma_N^e)^2 (E-\tilde{\vareps}_2)^2}.
\label{eq:T_F}
\end{equation}
Assuming a weak coupling between  $|f_1\rangle$ and $|f_2\rangle$ i.e. 
that 
$
t_c \ll 
\delta\vareps=|\vareps_+^e -\vareps_-^e|
$ 
(which also means that ${\vareps_-^e} \approx \tilde{\vareps}_2$),
for energies  close to  ${\vareps_-^e}$ 
the probability amplitude $T_e(E)$ can be further approximated by
\begin{equation}
 T_e(E)\approx \mathcal{A}
\frac{(\epsilon + q)^2}{\alpha^2 \epsilon^2 +1}
 \label{eq:Tfano}
\end{equation}
where $\mathcal{A}=\Gamma_N^e\Gamma_N^h/(\vareps_-^e-\tilde{\vareps}_2)^2$, $\epsilon=(E-\vareps_-^e)/\Gamma_N^e$, 
$\alpha^2=(\vareps_-^e-\vareps_+^e)^2/(\vareps_-^e-\tilde{\vareps}_2)^2$ 
and $q=(\vareps_-^e-\tilde{\vareps}_2)/ \Gamma_N^e$. 
Therefore, if  $0<\vareps_-^e,\tilde{\vareps}_2<\Delta$  
a  FR will appear in the subgap transport 
[see Fig.\ref{fig:theorycurves}(a)].  
For strong coupling such that 
$
\Gamma_N^e \gg |\vareps_-^e-\tilde{\vareps}_2|
$ 
the Fano lineshape would become a symmetric dip.  
If however $\vareps_+^e<\Delta < \vareps_-^e,\tilde{\vareps}_2$  is 
satisfied, a Breit-Wigner resonance (BWR)  of width ${\Gamma_N^e}$  
occurs [shown in Fig.~\ref{fig:theorycurves}(b)], while for $0<\vareps_+^e,\vareps_-^e,\tilde{\vareps}_2 <\Delta$
the Andreev reflection exhibits both a FR and a BWR 
[Fig.~\ref{fig:theorycurves}(c)]. 
Note, that $T_e(E)$ is very similar to the transmission 
amplitude $T_N(E)$ calculated in Ref.~\onlinecite{ref:thodoris} for 
normal conducting leads.
Since for $\sqrt{\Gamma_N^e \Gamma_N^h}\gtrsim |\sigma_S^{eh}|$
the resonance energies 
$\vareps_-^e$, $\vareps_+^e$ are usually very close to 
the resonance energies appearing in the expression of $T_N(E)$,
one finds that the resonance structures of the normal conductance 
will also appear in the sub-gap transport if the relevant 
resonance energies are smaller than the superconducting pair potential. 
This explains the occurrence of a Fano resonance in $G_{NS}(E)$ in 
Fig.~\ref{fig:trans2}. However, since $A(E)$ in 
Eq.~(\ref{eq:gns-finitE}) is usually much smaller than unity, 
$G_{NS}(E)$ itself can also be  smaller  than $G_N(E)$.  
The widths of the resonances in the Andreev-reflection coefficient 
can be significantly smaller than in the normal transmission. 
This happens because 
coupling to the superconductor does not lead to the broadening 
of the resonant levels. Therefore if 
$\sqrt{\Gamma_N^e \Gamma_N^h}\gg |\sigma_S^{eh}|$ the peaks in the 
normal and in the Andreev transport have roughly the same 
width while for $\sqrt{\Gamma_N^e \Gamma_N^h}\gtrsim |\sigma_S^{eh}|$
the width of the peaks in the Andreev reflection is  \emph{half}
of the width of the corresponding peaks in the normal transmission.
This can also be observed  in Fig.~\ref{fig:trans2}.
We note that for $E\approx\Delta$ where $\sigma_S^{eh}$ changes
rapidly with energy the formula shown in  
Eq.~(\ref{eq:gns-finitE}) is not applicable
because in the derivation  of Eq.~(\ref{eq:gns-finitE})  
we have assumed that the self energy $\sigma_S^{eh}$ is a slowly 
varying function of the 
energy.

Finally, we briefly discuss the predictions of our model for 
the  case when the coupling to the superconductor is stronger 
than to the normal lead, i.e. when 
$|\sigma_S^{eh}| \gtrsim \sqrt{\Gamma_N^e \Gamma_N^h}$.
The conductance can no longer be approximated  
by Eq.~(\ref{eq:gns-finitE}) because $\sigma_S^{eh}$ introduces 
hybridization between electron and hole levels. We find that 
in the most general case the conductance exhibits both a 
FR and a BWR, if the corresponding resonance energies are 
smaller than the superconducting gap.  
These peaks, as mentioned before, can be much narrower
than the ones in the normal transmission because  the 
superconductor does not broaden them. Moreover, we find that
for $|\sigma_S^{eh}| \gg \sqrt{\Gamma_N^e \Gamma_N^h}$ the 
conductance can even reach the unitarity limit. This could 
not happen in the opposite,  
$|\sigma_S^{eh}| \ll \sqrt{\Gamma_N^e \Gamma_N^h}$ case because
a resonance in $T_e(E)$  is not accompanied by a resonance
in $A(E)$  and therefore the conductance is always smaller than
$4e^2/h$. We illustrate this in Fig.~\ref{fig:theorycurves}(d)
where $G_{NS}$ is shown along with $G_N$. 
One can see that $G_N<2e^2/h$ because the 
couplings to the leads are asymmetric and there is a broad
resonance at $E/\Delta\approx 0.15$ along with an almost symmetric, 
 narrow dip at $E/\Delta\approx 0.32$. 
In contrast, $G_{NS}$  has a narrow FR and also a BWR, the latter
peak reaching the unitarity limit. 

In summary, we have studied the Andreev reflection through a class of 
molecules which exhibit Fano resonances in the normal conductance. 
Our numerical calculations based on \emph{ab initio} methods indicate that
Fano resonances may also appear in the sub-gap transport. 
A  simple theoretical model that we used to understand the results of  
the numerical calculations  predicts that a) 
if the coupling to the normal lead is weaker than the coupling 
 to the superconducting one, the 
resonance structure of the normal conductance can manifest itself
in the Andreev reflection coefficient 
if the resonance energies are smaller than 
the superconducting gap and  b)
if the coupling to the superconductor is strong, the resonances
in the normal conductance and in the Andreev reflection can be
very different, both in position and in width.

\section{Acknowledgment}
This work is supported partly by European Commission Contract
No.~MRTN-CT-2003-504574 and by EPSRC.

\appendix
\section{}
\label{sec:r_a-deriv}
There are numerous equivalent approaches 
to calculate transport coefficients through phase coherent normal-superconductor
hybrid systems\cite{ref:methods}. Here we employ the Green's function technique 
presented in Ref.~\onlinecite{ref:colin} in which the Hilbert space is divided into a sub-space $A$ containing
the external leads and a sub-space $B$ containing the molecule. 

Assuming for a moment that the molecule is isolated, for energies  
close to a resonance  
it can be described by quantum states $|f_1\rangle$, $|f_2\rangle$ with 
resonant energies $\vareps_1$, $\vareps_2$.  These states are coupled together
 by a hamiltonian $H_{12}$ with matrix element 
$t_c = \langle f_1\vert H_{12}\vert f_2\rangle$.
The effect of coupling of the molecule to the normal conducting  
(superconducting) lead via a coupling matrix $W_N$ ($W_S$) 
is represented by the energy dependent 
self-energy matrices 
$\mbox{\boldmath $\Sigma$}_N=\mbox{\boldmath $\sigma$}_N-i\mbox{\boldmath $\Gamma$}_N$ 
($\mbox{\boldmath $\Sigma$}_S=\mbox{\boldmath $\sigma$}_S-i\mbox{\boldmath $\Gamma$}_S$) 
where $\mbox{\boldmath $\sigma$}_N$, $\mbox{\boldmath $\Gamma$}_N$ 
($\mbox{\boldmath $\sigma$}_S$, $\mbox{\boldmath $\Gamma$}_S$) are hermitian. 
We assume that the coupling matrices are diagonal in the quasiparticle 
$e,h$ space:
\begin{equation}
W_{N,S}=
\left(%
 \begin{array}{cc}
   W_{N,S}^e & 0 \\
   0 & W_{N,S}^h \\
 \end{array}%
 \right),\\
 \end{equation}
where $W_{N\,(S)}^e=-(W_{N\,(S)}^h)^*$. Since the Green's function of the 
(isolated) normal lead is also diagonal in the quasiparticle space, so will be 
$\mbox{\boldmath $\Sigma$}_N=
\textnormal{Diag}(\mbox{\boldmath $\sigma$}_{N}^{e}
-i\mbox{\boldmath $\Gamma$}_N^{e},
\mbox{\boldmath $\sigma$}_{N}^{h}-i\mbox{\boldmath $\Gamma$}_N^{h})
$, too. 
The self energy coming from the coupling to the superconductor has both diagonal 
and off-diagonal parts, but for $E\le\Delta$ 
it reads
\begin{equation}
\mbox{\boldmath $\Sigma$}_S =\left(%
 \begin{array}{cc}
   \mbox{\boldmath $\sigma$}_S^{e} &  \mbox{\boldmath $\sigma$}_S^{eh}\\
    \mbox{\boldmath $\sigma$}_S^{he}& \mbox{\boldmath $\sigma$}_S^{h} \\
 \end{array}%
 \right)\\
\end{equation}
i.e. the superconducting lead does not broaden the levels. Moreover, since 
$|f_2\rangle$ is only coupled with $|f_1\rangle$
but not with any of the leads, the self-energy matrix elements 
of the matrices $\mbox{\boldmath $\Sigma$}_N$, 
$\mbox{\boldmath $\Sigma$}_S$  will only affect 
the resonance energy $\vareps_1$ of the backbone state but not the 
energy $\tilde{\vareps}_2=\vareps_2-E_F$ of the side coupled state.
We now introduce the following notations:  
$\tilde{\vareps}_1^{e,h}=\vareps_1-E_F-({\sigma}_N^{e,h}+{\sigma}_S^{e,h})$ 
[where $\tilde{\sigma}_{N,S}^{e,h}$ are the (only) nonzero element of the 
matrices  $\mbox{\boldmath $\sigma$}_{N,S}^{e,h}$],
$\bar{\vareps}^{\,e,h}=(\tilde{\vareps}_1^{\,e,h}+\tilde{\vareps}_2)/2$, 
$\delta{\vareps}^{\,e,h}=(\tilde{\vareps}_1^{\,e,h}-\tilde{\vareps}_2)/2$, 
$\vareps_{\pm}^{e,h}=\bar{\vareps}^{\,e,h}\pm\sqrt{(\delta\vareps^{\,e,h})^2+t_c^2}$. 
Denoting by $\mathbf{G}_{BB}(E)$ the retarded  Green's function of the molecule 
and using the formula\cite{ref:colin} 
\begin{equation}
R_A=\mbox{Tr}[\mbox{\boldmath $\Gamma$}_N^{e}\mathbf{G}_{BB}(E)\mbox{\boldmath $\Gamma$}_N^{h}\mathbf{G}^{\dagger}_{BB}(E)]
\end{equation}
to calculate the probability of the Andreev reflection, we find  
after straightforward calculations that
\begin{equation}
R_A= 
\frac{4 \Gamma_N^e\Gamma_N^h(E-\tilde{\vareps}_2)^2(E+\tilde{\vareps}_2)^2 (\sigma_S^{eh})^2}{|D|^2}.
\end{equation}
Here the denominator is 
\begin{widetext}
\begin{equation}
D=[(E-\vareps_{+}^{e})(E-\vareps_{-}^{e})+ i (E-\tilde{\vareps}_2)\Gamma_N^e]
[(E+\vareps_{+}^{h})(E+\vareps_{-}^{h})+ i (E+\tilde{\vareps}_2)\Gamma_N^h]
-(\sigma_S^{eh})^2(E-\tilde{\vareps}_2)(E+\tilde{\vareps}_2).
\end{equation}
\end{widetext}
and $\sigma_S^{eh}$, $\Gamma_N^e$, $\Gamma_N^h$ are 
the only non-zero elements of the matrices 
$\mbox{\boldmath $\sigma$}_S^{eh}=\mbox{\boldmath $\sigma$}_S^{he}$, 
$\mbox{\boldmath $\Gamma$}_N^{e}$, $\mbox{\boldmath $\Gamma$}_N^{h}$.

\end{document}